\documentclass[12pt,preprint]{aastex}
\makeatletter
\renewcommand{\fnum@figure}{Figure~\thefigure}
\makeatother
\usepackage{epsfig}
\usepackage{natbib}
\usepackage{graphicx}
\usepackage{slashbox}
\usepackage{multirow}
\usepackage{lscape}
\usepackage{mathrsfs,amssymb}
\usepackage{amsmath}
\usepackage{cite}
\usepackage{float}
\usepackage[section]{placeins}
\usepackage{booktabs}
\let\captionbox\relax
\usepackage{caption}
\usepackage{enumitem}
\usepackage{lineno}

\newcommand       \kpc          {\,{\rm kpc}}

\newcommand       \simlt        {\lesssim}
\newcommand       \simgt        {\gtrsim}

\newcommand       \mum          {\,{\rm \mu m}}

\newcommand       \simali       {\sim\,}

\newcommand       \NCaro      {N_{\rm C,aro}}
\newcommand       \NCali      {N_{\rm C,ali}}

\newcommand       \alifrac      {\eta_{\rm ali}}
\newcommand       \Iratioobs   {\left(P_{3.4}/P_{3.3}\right)_{\rm obs}}
\countdef\decade=200
\advance\decade by \year
\countdef\hours=201
\advance\hours by \time
\divide\hours by 60
\countdef\mins=202
\advance\mins by \hours
\multiply\mins by 60
\multiply\hours by 100
\countdef\miltime=203
\advance\miltime by \hours
\advance\miltime by \time
\advance\miltime by -\mins
\def\today{\number\decade.\number\month.\number\day.\number\miltime}

\shorttitle{The Extinction Curve toward HD\,93222}
\title{
\vspace*{-2.0em}
{\normalsize\rm Accepted for publication in
               {\it The Astrophysical Journal Letters}}\\
\vspace*{1.0em}
Widespread Detection of Aromatic and Aliphatic Emission
in the Dual Quasar J0749+2255 at Cosmic Noon
\\{\small DRAFT: \today ~~}
}
\author{C.~E.~Mentzer\altaffilmark{1},
            Aigen Li\altaffilmark{1},
            and X.J.~Yang\altaffilmark{2}}
\altaffiltext{1}{Department of Physics and Astronomy,
                  University of Missouri,
                  Columbia, MO 65211, USA
                  {\sf lia@missouri.edu}}        
\altaffiltext{2}{Department of Physics,
                      Xiangtan University,
                      411105 Xiangtan, Hunan Province, China;
                      {\sf xjyang@xtu.edu.cn}}
\begin{document}

\begin{abstract}
Based on the integral field observations
made with the {\it Mid Infrared Instrument} (MIRI) 
aboard the {\it James Webb Space Telescope} (JWST),
we report a widespread detection of aromatic and
aliphatic hydrocarbon emission at rest-frame
3.3 and 3.4$\mum$ in SDSS\,J074922.96+225511.7
(hereafter J0749+2255),
a dual quasar at redshift $z$\,$\approx$\,2.17,
corresponding to a cosmic age of
$\simali$3 billion years after the Big Bang,
a time period known as the ``cosmic noon''
when star formation and black hole growth peak.
%
With the 3.3$\mum$ emission ascribed to
aromatic C--H stretches of small polycyclic
aromatic hydrocarbon (PAH) molecules
and the 3.4$\mum$ emission assigned
to aliphatic C--H stretches of aliphatic
sidegroups attached to PAHs,
we utilize the observed intensities of 
the 3.3 and 3.4$\mum$ emission features
to estimate the aliphatic fractions of PAHs
and their spatial variations across J0749+2255,
which is, to our knowledge, the most distant
object to date in which both aromatics and
aliphatics have ever been detected.
We find that both the 3.3 and 3.4$\mum$
emission features are pronounced and the aliphatic
fractions are surprisingly high in the most luminous
regions centered on the two quasar nuclei,
suggesting that not only small PAHs
(of $\simali$20--30 carbon atoms)
but also their attached aliphatic sidegroups 
survive in extreme starburst activities
and quasar-driven shocks.
\end{abstract}

\section{Introduction\label{sec:intro}}
The broad 3.3$\mum$ emission feature, first detected
in NGC~7027, a planetary nebula  (Merrill et al.\ 1975),
is attributed to the {\it aromatic} C--H stretching
vibrations of small, neutral polycyclic aromatic
hydrocarbon (PAH) molecules of several tens of
carbon atoms (L\'eger \& Puget 1984,
Allamandola et al.\ 1985). This feature was recently
detected by the {\it Mid Infrared Instrument} (MIRI) 
on board the {\it James Webb Space Telescope} (JWST)
in SPT0418-47, a gravitationally-lensed galaxy
at a redshift of $z$\,$\approx$\,4.22,
when the Universe was only $\simali$11\%
of its current age (Spilker et al.\ 2023). 
More recently, Chen et al.\ (2024) reported
the detection of this feature also by JWST/MIRI
in SDSS\,J074922.96+225511.7
(hereafter J0749+2255),
a dual quasar at redshift $z$\,$\approx$\,2.17.
The 3.3$\mum$ aromatic C--H emission,
as bright as $\simali$1\% of the infrared (IR)
luminosity at 8--1000$\mum$,
has also been seen in 32 (of 37)  star-forming
galaxies and dust-obscured active galactic nuclei
(AGNs) at $z$\,$\approx$\,0.65--2.46
(McKinney et al.\ 2026).
%

The 3.3$\mum$ aromatic feature is sometimes
accompanized by a weak satellite emission
feature at 3.4$\mum$ arising from {\it aliphatic}
C--H stretches (see Yang et al.\ 2017,
Allamandola et al.\ 2021, Boersma et al.\ 2023;
but also see Tokunaga \& Bernstein 2021,
Tokunaga et al.\ 2025).
This indicates that astronomical PAHs
may have an {\it aliphatic} component,
either in the form of aliphatic sidegroups
like methyl (--CH$_3$) attached as
functional groups, or in the form of
superhydrogenation (i.e., excess H atoms
are added to peripheral C atoms and
convert the originally aromatic C--H bonds
into aliphatic bonds;
see Bernstein et al.\ 1996, Yang et al.\ 2020).
The intensity ratio of the 3.4$\mum$ feature
to the 3.3$\mum$ feature allows one to
quantitatively probe the aliphatic contents
of the emitter (Li \& Draine 2012, Yang \& Li 2023).

As the aromatic and aliphatic C--H stretches
at 3.3 and 3.4$\mum$ are often much weaker
than the aromatic C--C stretches and C--H
bending modes of PAHs at 6.2, 7.7, 8.6, and
11.3$\mum$, the simultaneous detection of
both the 3.3 and 3.4$\mum$ features in extragalactic
objects in the pre-JWST era was limited to
a handful of nearby, bright galaxie
(e.g., see Sturm et al.\ 2000, Imanishi et al.\ 2010,
Kondo et al.\ 2012, Lee et al.\ 2012,
Yamagishi et al.\ 2012, Inami et al.\ 2018,
Lai et al.\ 2020). 
Thanks to its unprecedented sensitivity
and wavelength coverage,
the rest-frame 3.3 and 3.4$\mum$ features
were recently detected in several dozens
of star-forming galaxies at redshifts
$z$\,$\simali$0.2--0.5, based on observations
made with the {\it Wide-Field Slitless Spectroscopy}
(WFSS) mode of the {\it Near-Infrared Camera}
(NIRCam) on board JWST (Lyu et al.\ 2025).

In this work, we analyze the JWST/MIRI
{\it Integral Field Unit} (IFU) spectra of
the dual quasar J0749+2255
at $z$\,$\approx$\,2.17 and find that,
in addition to the 3.3$\mum$ emission,
the 3.4$\mum$ emission is also widespread
in this dual quasar system.
This marks the detection
of both aromatics and aliphatics
in the early Universe at a cosmic age of
$\simali$3 billion years after the Big Bang,
a time period known as the ``cosmic noon'',
which is believed to be the peak epoch
of star formation and black hole growth.
We note that in SPT0418-47, one potential
secondary feature was seen at rest-frame
3.4$\mum$, but the lack of spectral coverage
redward of this feature made its detection
more tentative because the underlying continuum
was poorly constrained (Spilker et al.\ 2023).
Therefore, to our knowledge, J0749+2255 is
the most distant galaxy to date in which both
aromatics and aliphatics have ever been detected.
Prior to this, GN-IRS-58, a star-forming galaxy
at $z$\,$\approx$\,1.998, has been observed
by JWST/MIRI to show both aromatic and aliphatic
emission (McKinney et al.\ 2026).
  

%
This paper is organized as follows.
In \S\ref{sec:data} we briefly describe the target
J0749+2255 and the JWST/MIRI data extraction procedure.
We present in \S\ref{sec:spectra}
the PAH emission spectra across J0749+2255
retrieved from the JWST/MIRI IFU observations.
In \S\ref{sec:discussion} we derive the aliphatic fractions
of PAHs and discuss their implications.
Our major results are summarized
in \S\ref{sec:summary}. 
This double quasar at cosmic noon
is of particular interest for the (co-)evolution
of galaxies and super-massive black holes
(SMBHs) and for the growth of SMBHs in general.
The findings derived in this study demonstrate
some open questions in the current understanding
of PAH astrophysics, and urge caution
in the interpretation of future observations of this kind.
They also suggest a potential new diagnostic
value in identifying complex structures
and morphology in such objects.

\section{The Target and the Data}\label{sec:data}
Dual quasars are rare astronomical phenomena
where two quasars are found in close proximity,
typically within interacting or merging galaxies.
The presence of dual quasars suggests
a scenario where two galaxies, each hosting
a SMBH, are merging or interacting closely.
This interaction leads to intense gravitational forces,
causing the black holes to feed on
surrounding material and shine brightly as quasars.
Therefore, dual quasars provide valuable insights
into the dynamics of galaxy mergers,
the growth of SMBHs, and the evolution
of galaxies over cosmic time.
As PAH emission has been widely used
as diagnostics of the relative roles of
starbursts and AGNs in galaxies
(e.g., see Genzel et al.\ 1998),
dual quasars provide a valuable opportunity
to explore the effects of quasar-driven
shocks and galaxy-merging-induced
starbursts on the excitation and destruction
of PAH molecules.

\begin{figure*}[htp]
\vspace{-1mm}
\begin{center}
\includegraphics[width=11.2cm,angle=0]{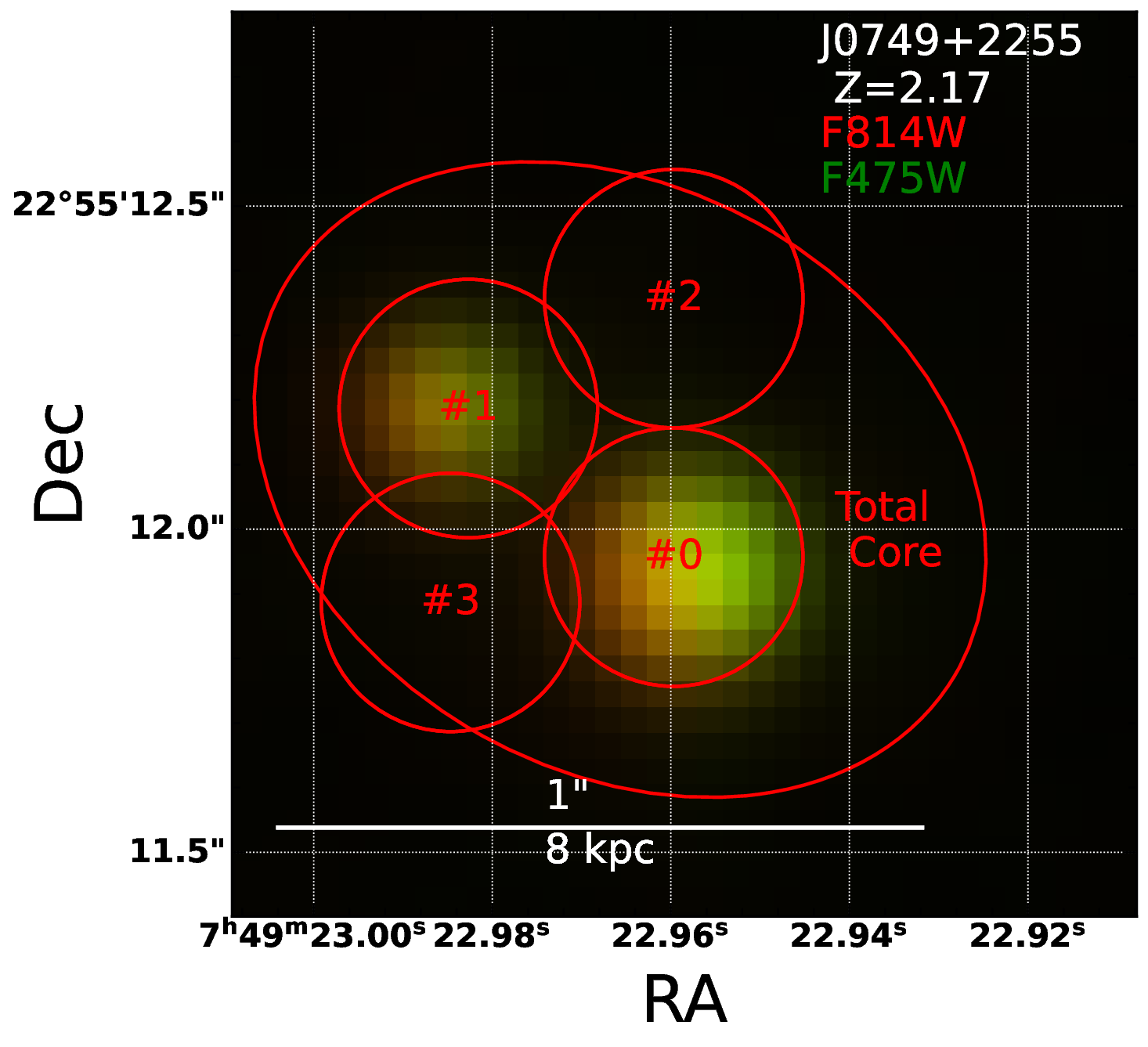}
\end{center}
\vspace{-4mm}
\caption{\label{fig:hst}\footnotesize
Color composite image of the dual qusar
J0749+2255 at $z$\,$\approx$\,2.17,
created from observations obtained with
HST/WFC3's F475W (green) and F814W (red) filters. 
Superimposed are two small circles (\#0, \#1)
centered on the quasar cores and two offset regions
(\#2, \#3) from these cores.
The larger ellipse contains
the central brightest region
(see Figure~\ref{fig:miri}b).
Each of the small circles has an aperture
size of 1.7$\kpc$, while the size of
the large ellipse is 3.7$\kpc$\,$\times$\,5.0$\kpc$.
         }
\vspace{-3mm}
\end{figure*}

J0749+2255 is the first kpc, distant dual quasar
with a projected separation of $\simali$3.8$\kpc$
(Chen et al.\ 2023). As shown in Figure~\ref{fig:hst},
the separation of two distinct quasar cores
within this merging system is clearly seen
in the color composite image of J0749+2255,
created from observations obtained with
the {\it Wide Field Camera 3} (WFC3) on board
the {\it Hubble Space Telescope} (HST),
utilizing the F475W and F814W filters.

As a part of the JWST Cycle 1 GO Proposal \#2654
``Kpc-scale dual supermassive black holes
and their impact on galaxy formation
at cosmic noon'' (PI Yuzo Ishikawa),
spatially-resolved infrared (IR) spectra 
of J0749+2255 were obtained
using the {\it Medium Resolution Spectrometer}
(MRS) mode of JWST's MIRI instrument
on November 21--22, 2022.
Observations were conducted in long grating mode
across multiple channels. For this work, we use
the second-channel wavelength range of
10.02–-11.70$\mum$, which contains
the rest-frame 3.3 and 3.4$\mum$ aromatic
and aliphatic emission features.
All data used in this study are publicly available
through the {\it Mikulski Archive for Space Telescopes}
(MAST). Data reduction is performed
using the JWST calibration pipeline version 1.11.4
(Chen et al.\ 2024), along with the calibration CRDS
version 11.17.6 with context {\tt jwst$_{-}$1193.pmap}.
Spectra are extracted from the cubes
through aperture photometry
using {\tt Photutils} (Bradley et al.\ 2024). 

\begin{figure*}[htp]
\vspace{-1mm}
\begin{center}
\includegraphics[width=11.2cm,angle=0]{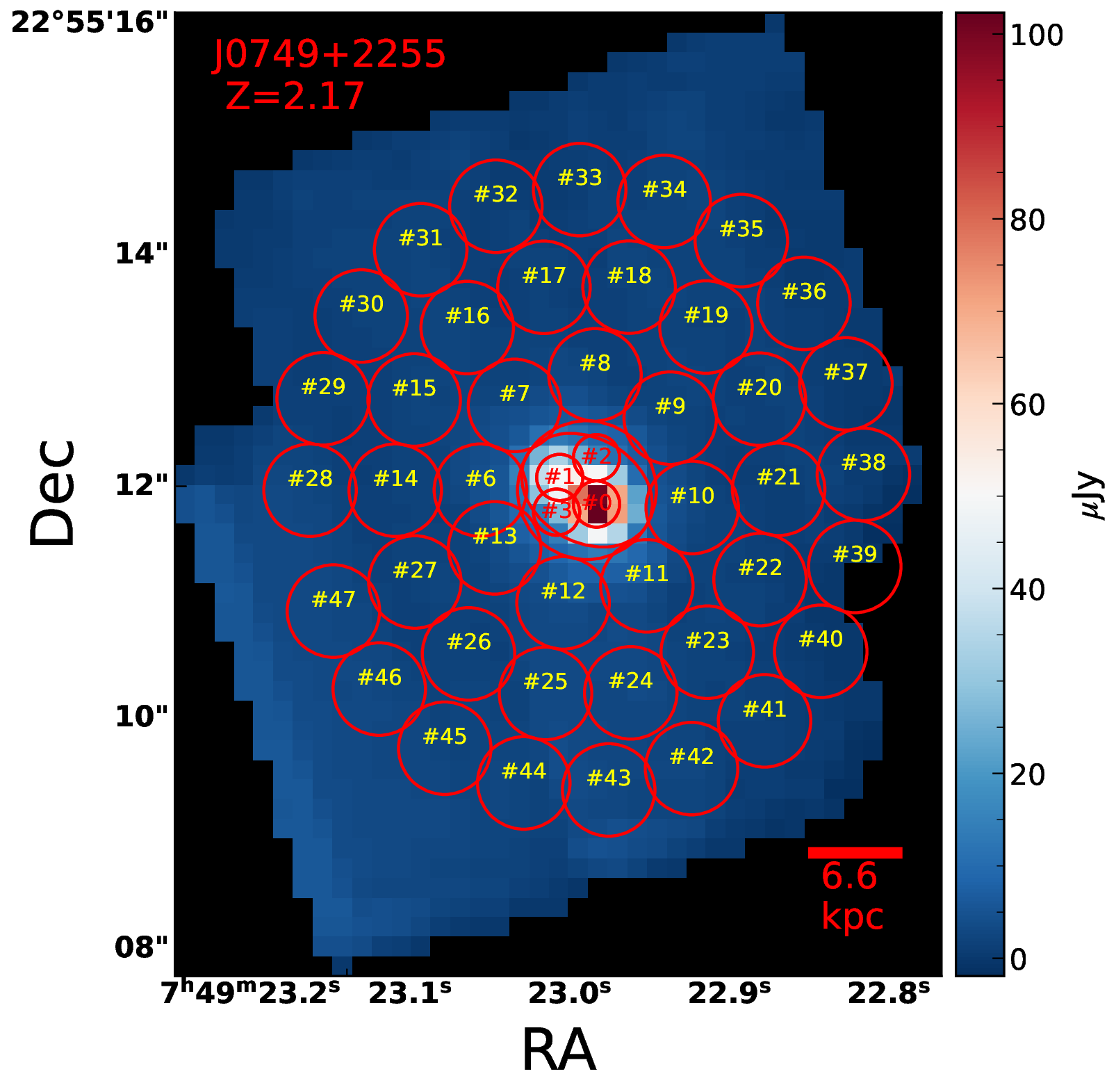}\\
\includegraphics[width=11.2cm,angle=0]{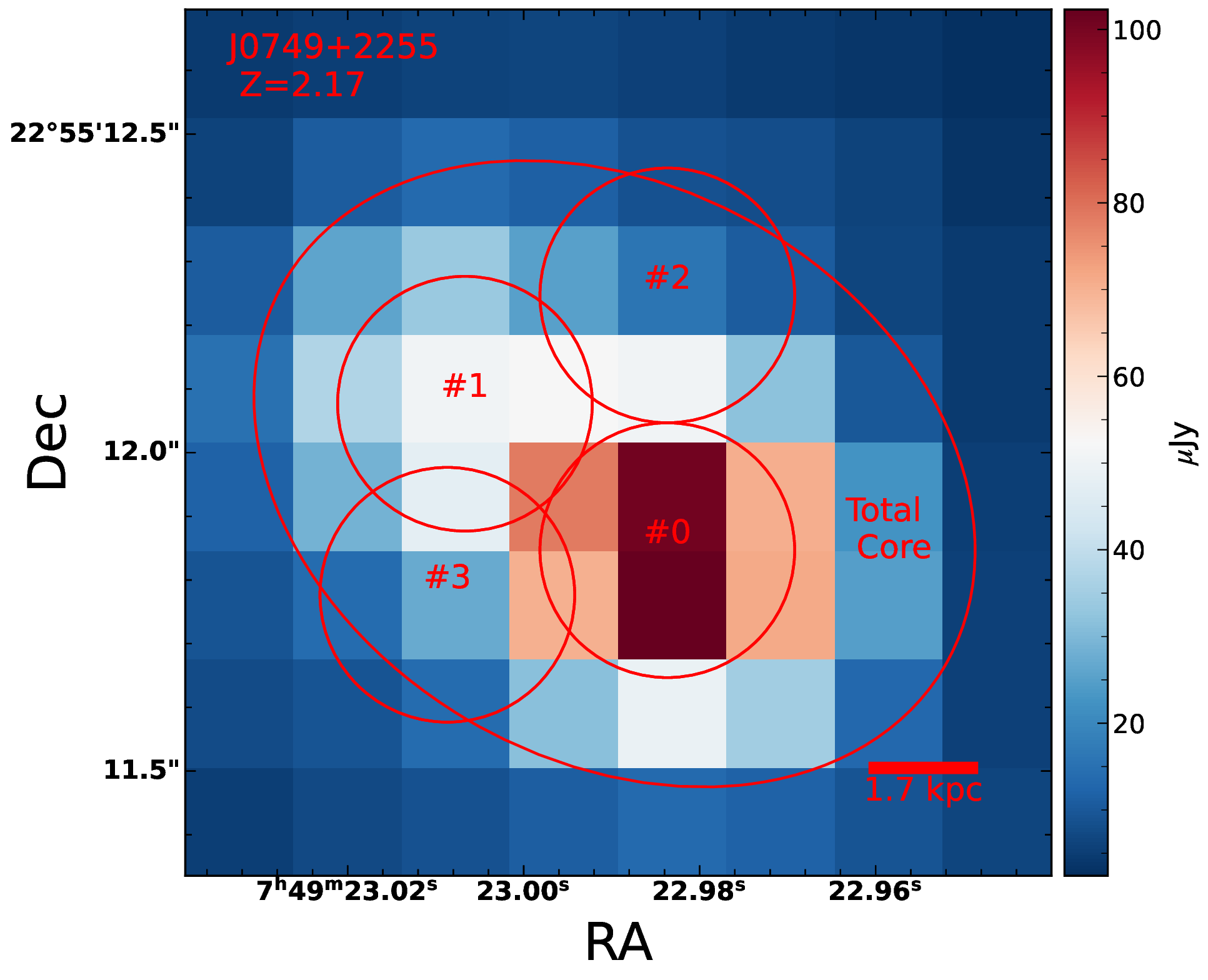}
\end{center}
\vspace{-4mm}
\caption{\label{fig:miri}\footnotesize
JWST's MIRI/MRS images of the entire region
(upper panel, a)
and the zoomed-in brightest center
(bottom panel, b)
of the dual quasar system J0749+2255.
For spectral extraction,
we take 42 large circular apertures
(with 6.6$\kpc$ for each aperture)
for the outer region surrounding
the quasar cores (from \#6 to \#47),
and four small circular apertures
(with 1.7$\kpc$ for each aperture)
for the bright central region (from \#0 to \#3).
In addition, a large ellipse (aperture \#5)
for the entire inner region is also taken.
The small circles and the large ellipse have
the same size as that in Figure~\ref{fig:hst}.
         }
\vspace{-3mm}
\end{figure*}

To analyze the spatially-resolved JWST/MIRI
spectra of J0749+2255, as illustrated in
Figure~\ref{fig:miri}a, we take 42 apertures
surrounding the quasar cores, from \#6 to \#47,
with each aperture spanning a diameter of 6.6$\kpc$.
We refer the region covered by these surrounding
apertures the ``outer region''.
Given the distance of this dual quasar
and the resulting angular size from
$z$\,$\approx$\,2.17, smaller apertures than this
would fall below the artifact threshold
rendering the already noisy data unusable.
For the ``outer region'', we used a tiling effect
to effectively sample the entire region,
while maintaining a usable signal-to-noise ratio
(SNR), arranging circles around the center
in three distinct features.

In contrast, the central region is significantly
brighter, allowing for smaller apertures to be taken,
while maintaining a usable SNR
for the purposes of detecting PAH emission,
both aliphatic and aromatic.
Therefore, as zoomed in Figure~\ref{fig:miri}b, 
we take four smaller apertures
(with 1.7$\kpc$ for each) within the bright
central region, from \#0 to \#3:
one aperture on each core (\#0, \#1), and
two offset apertures from these cores (\#2, \#3).
In addition, we also take a big,
3.7$\kpc$\,$\times$\,5.0$\kpc$
elliptical aperture (\#5) which encompasses
the entirety of the inner region
(which includes all the four small apertures).
In the following, we refer the bright central region
(measured by the four small circles
and the large ellipse) the ``inner region''.
We have also attempted a pixel-by-pixel analysis.
However, the SNR from this was too low,
making aperture photometry the most appropriate
method to extract a 2-dimensional aromatic
and aliphatic emission landscape. 

To summarize, as illustrated in
Figure~\ref{fig:miri}, a zeroth-order
moment map of the second MIRI/MRS
channel (which shows the entire region
of this dual quasar system
along with the close-up inner region),
we extract a total of 47 spectra,
with 42 spectra for the outer apertures
and five spectra for the inner apertures.
We will utilize these spectra in next sections
for spectral decomposition
and alphatic fraction determination.

\section{Widespread Detection of Aromatic
           and Aliphatic Hydrocarbon Emission}\label{sec:spectra}
\begin{figure*}[htp]
\vspace{-2mm}
\begin{center}
\includegraphics[width=13.6cm,angle=0]{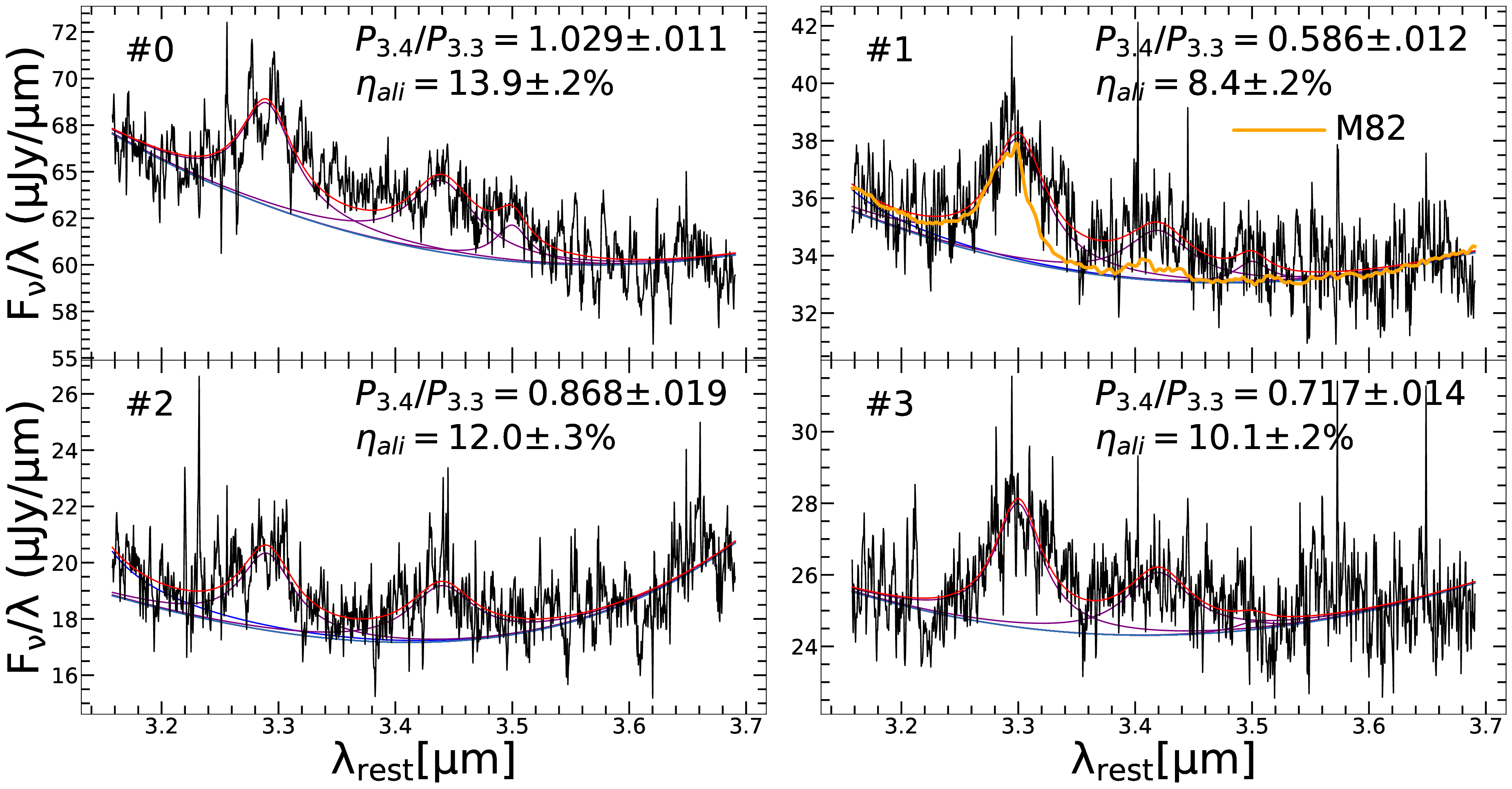}\\
\includegraphics[width=13.6cm,angle=0]{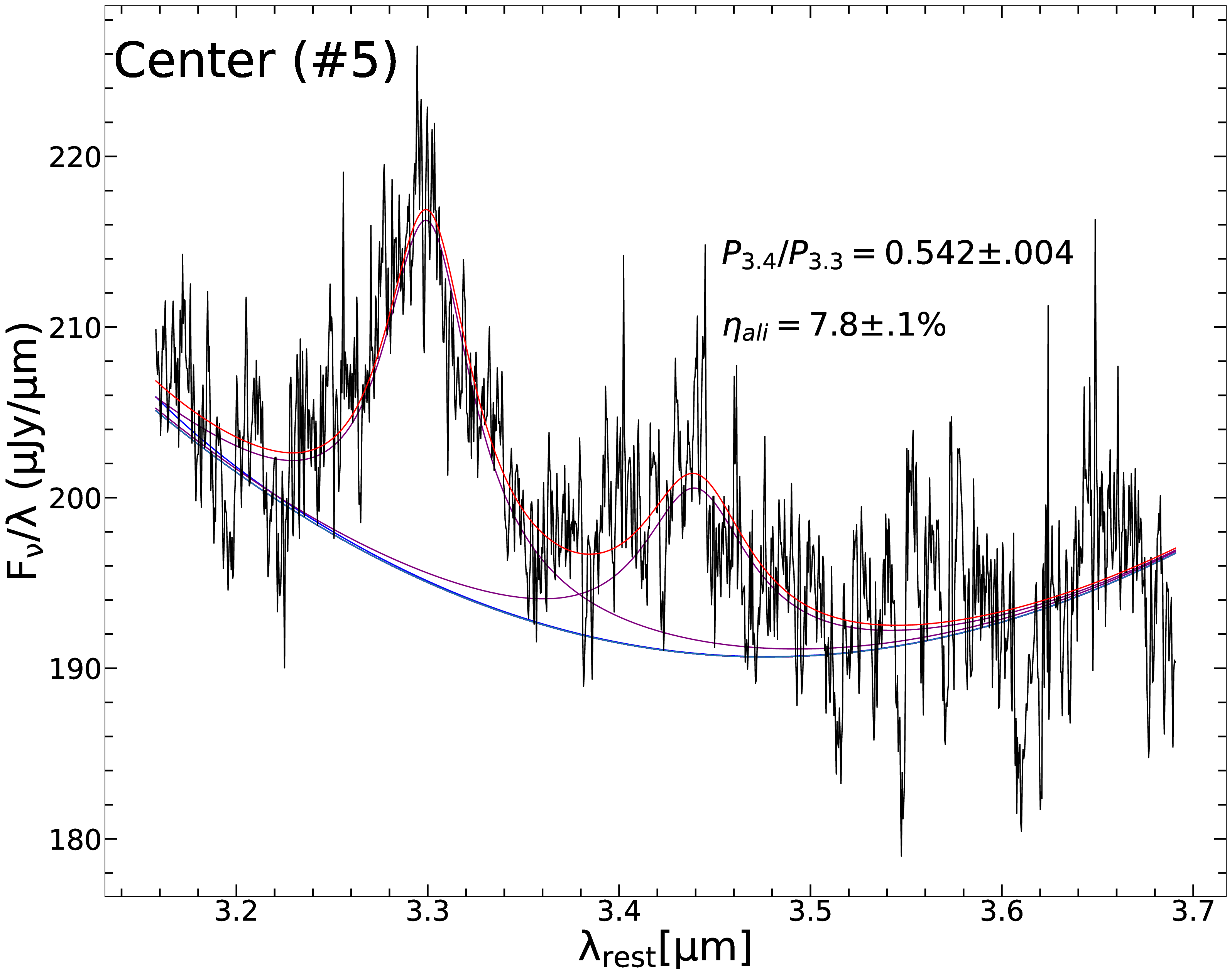}
\end{center}
\vspace{-7mm}
\caption{\label{fig:spectra_inner}\footnotesize
JWST/MIRI rest-frame spectra extracted for
the brightest inner regions of J0749+2255:
two small apertures of 1.7$\kpc$ each
for the quasar cores (\#0, \#1) and two apertures
offset from the cores (\#2, \#3),
as well as the entire inner
region (\#5) represented by a large
3.7$\kpc$\,$\times$\,5.0$\kpc$ ellipse
(see Figure~\ref{fig:hst} and Figure~\ref{fig:miri}b).
Utilizing the PAHFIT tool, the observed spectra
are decomposed into two Drude profiles
(magenta lines) and an underlying continuum
(blue line), with the two Drude profiles characterizing
the 3.3$\mum$ aromatic and 3.4$\mum$
aliphatic features, and the red line as the sum of
the two Drude profiles and the continuum.
For Regions \#0 and \#1, a secondary subfeature
at $\simali$3.49--3.51$\mum$ is also shown.
For comparison, we also overlay
the JWST/NIRSpec spectrum of
the starburst galaxy M82
(orange line; Sturm et al.\ 2000)
to that of \#1, after normalized to
the 3.3$\mum$ peak of \#1.
Clearly, the PAH molecules in J0749+2255
are more aliphatic than those in M82.
In each panel, the ratio of the power emitted
in the 3.4$\mum$ feature to that in the 3.3$\mum$
feature ($P_{3.4}/P_{3.3}$) is labelled. Also labelled
is $\alifrac$, the aliphatic fraction of PAHs
(see \S\ref{sec:discussion}).
         }
\vspace{-3mm}
\end{figure*}

As elaborated in \S\ref{sec:data}, 
the central region, defined as the brightest inner area
of the dual quasar system, is analyzed using 1.7$\kpc$
apertures to encase each quasar core (i.e., \#0, \#1),
followed by offsetting the apertures to investigate
the adjacent regions (i.e., \#2, \#3).
In addition, an elliptical aperture (i.e., \#5)
surrounding the entire brightest central region
is used to examine the center as a whole.
We extract the MIRI spectra of these regions
and show in Figure~\ref{fig:spectra_inner}.
It is apparent that, within these apertures,
the rest-frame 3.3 and 3.4$\mum$ features
are seen clearly, which are attributed to
aromatic and aliphatic C--H stretches,
respectively (see Yang et al.\ 2017).
A close inspection of the 3.4$\mum$ feature
sometimes also reveals the presence of one
or more subfeatures at 3.43, 3.46, 3.51,
and 3.56$\mum$. 
These weak subfeatures
have been observed in various sources,
both in the pre-JWST era
(e.g., see Geballe et al.\ 1985,
Jourdain de Muizon et al.\ 1986,
Joblin et al.\ 1996) 
and in the JWST era (e.g., see Lai et al.\ 2023,
Chown et al.\ 2024,
Schroetter et al.\ 2024,
van de Putte et al.\ 2025).
They are often attributed to
aliphatic functional groups
such as CH$_3$ or CH$_2$
(see Yang et al.\ 2017)
and/or the anharmonicity
of the aromatic C--H stretch
(Barker et al.\ 1987,
Esposito et al.\ 2024).
As H$_2$ also emits at $\simali$3.438,
3.485, and 3.500$\mum$
(e.g., see Figure~1 of Thatte et al.\ 2026),
caution need to be taken
when assigning these subfeatures.

For regions \# 0 and \# 1, directly on the quasars,
the 3.3$\mum$ emission is noticeably more intense
than that in the offset regions. This is unanticipated 
as one may think that small PAHs would have been
destroyed in the core regions where ultraviolet (UV)
photons are rich and shocks may be generated.
Furthermore, the relative strength
of the 3.4$\mum$ aliphatic C--H feature
(with respect to the 3.3$\mum$ aromatic C--H feature)
does not appear to show systematic variations
from the cores to the offset regions.
This is also unexpected as one may think that
the aliphatic sidegroups attached to PAHs
(which give rise to the 3.4$\mum$ emission)
could have been ruptured in the UV-rich quasar cores.
This will be discussed in \S\ref{sec:discussion}
more quantitatively.

The surrounding environment, defined as a large
circular region out to the edge of Channel~2 field
of view, is next examined. This region is of particular
interest as it did not initially appear to have anything
of note upon the initial inspection of the data.
However, after a closer examination,
this ``uninteresting'' region also shows
widespread 3.3 and 3.4$\mum$ emission.

To examine this ``outer'' region, larger apertures of
roughly 6.6$\kpc$ are used in an attempt to
improve the SNR, given that the emission is
only a few microjanksy.
%
%
As illustrated in Figure~\ref{fig:miri}a,
the so-called surrounding ``outer region''
(distinguished from the brighest inner region)
is divided into three bands:
the inner band (\#6, \#7, ..., \#13),
the intermediate band (\#14, \#15, ..., \#27),
and the outermost band
(\#28, \#29, ..., \#47).

We retrieve the JWST/MIRI spectra for
each of the apertures via aperture photometry,
and show in Figure~\ref{fig:spectra_outer}a,\,b,\,c
for the apertures in the inner band, intermediate band
and outermost band, respectively.
The spectra of about half (20/42) of these apertures
exhibit prominent emission features at 3.3 and
3.4$\mum$ in the rest frame.
Both the intensities of these features and their relative
strengths show considerable regional variations
(see \S\ref{sec:discussion} for quantitative discussions).


\begin{figure*}[htp]
\begin{center}
\includegraphics[width=15.0cm,angle=0]{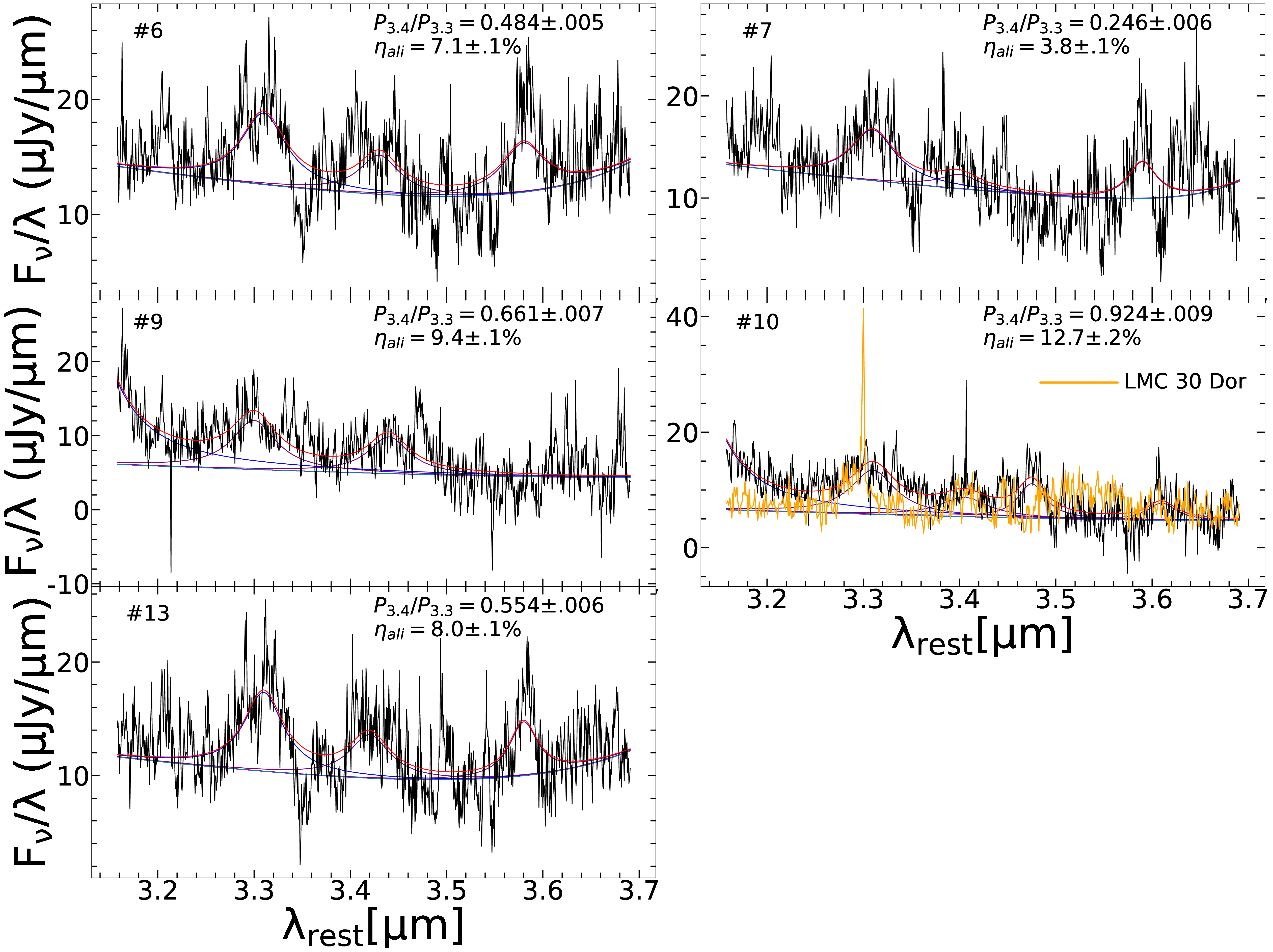}
\end{center}
\vspace{-4mm}
\caption{\label{fig:spectra_outer}\footnotesize
JWST/MIRI rest-frame spectra extracted for
the inner band surrounding the brightest
center of J0749+2255 (see Figure~\ref{fig:miri}a).
Each circle represents an aperture of 6.6$\kpc$.
Similar to Figure~\ref{fig:spectra_inner},
the observed spectra are decomposed
with PAHFIT into two Drude profiles
(magenta lines) and an underlying continuum
(blue line), and in some cases, secondary subfeatures
at $\simali$3.xx--3.xx$\mum$ are also included.
For comparison, we also overlay
the JWST/NIRSpec spectrum of
the 30 Doradus star-forming complex
in the Large Magellanic Cloud
(orange line; Sturm et al.\ 2000)
to that of \#10, after normalized to
the 3.3$\mum$ peak of \#10.
  }
\vspace{-3mm}
\end{figure*}

\begin{figure*}[htp]
\figurenum{\ref{fig:spectra_outer}}
\leavevmode
\begin{center}
\includegraphics[width=15.0cm,angle=0]{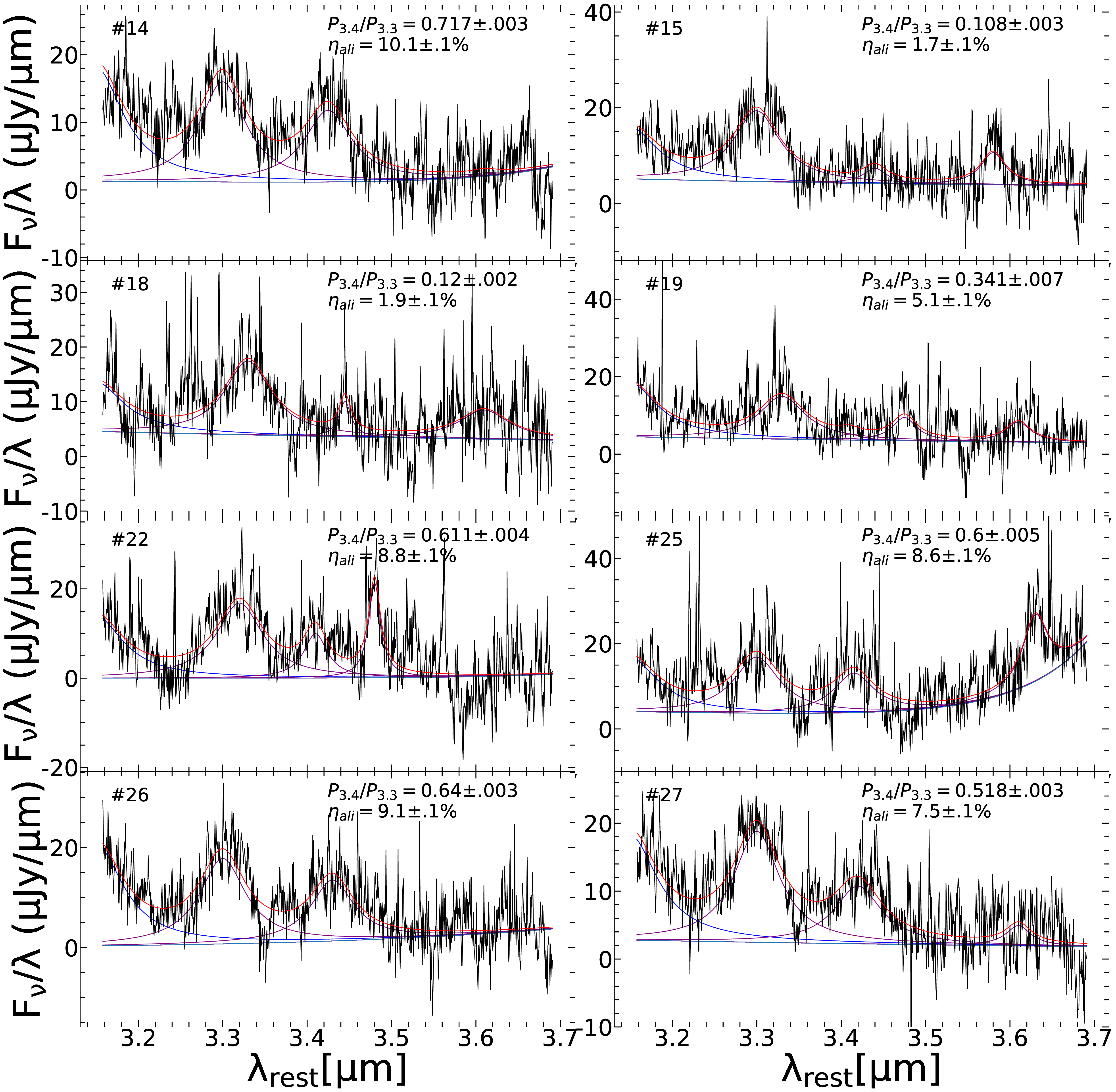}
\end{center}
\vspace{-4mm}
\caption{\footnotesize
  Continued, but for the intermediate band.
  }
\vspace{-3mm}
\end{figure*}

\begin{figure*}[htp]
\figurenum{\ref{fig:spectra_outer}}
\leavevmode
\begin{center}
\includegraphics[width=15.0cm,angle=0]{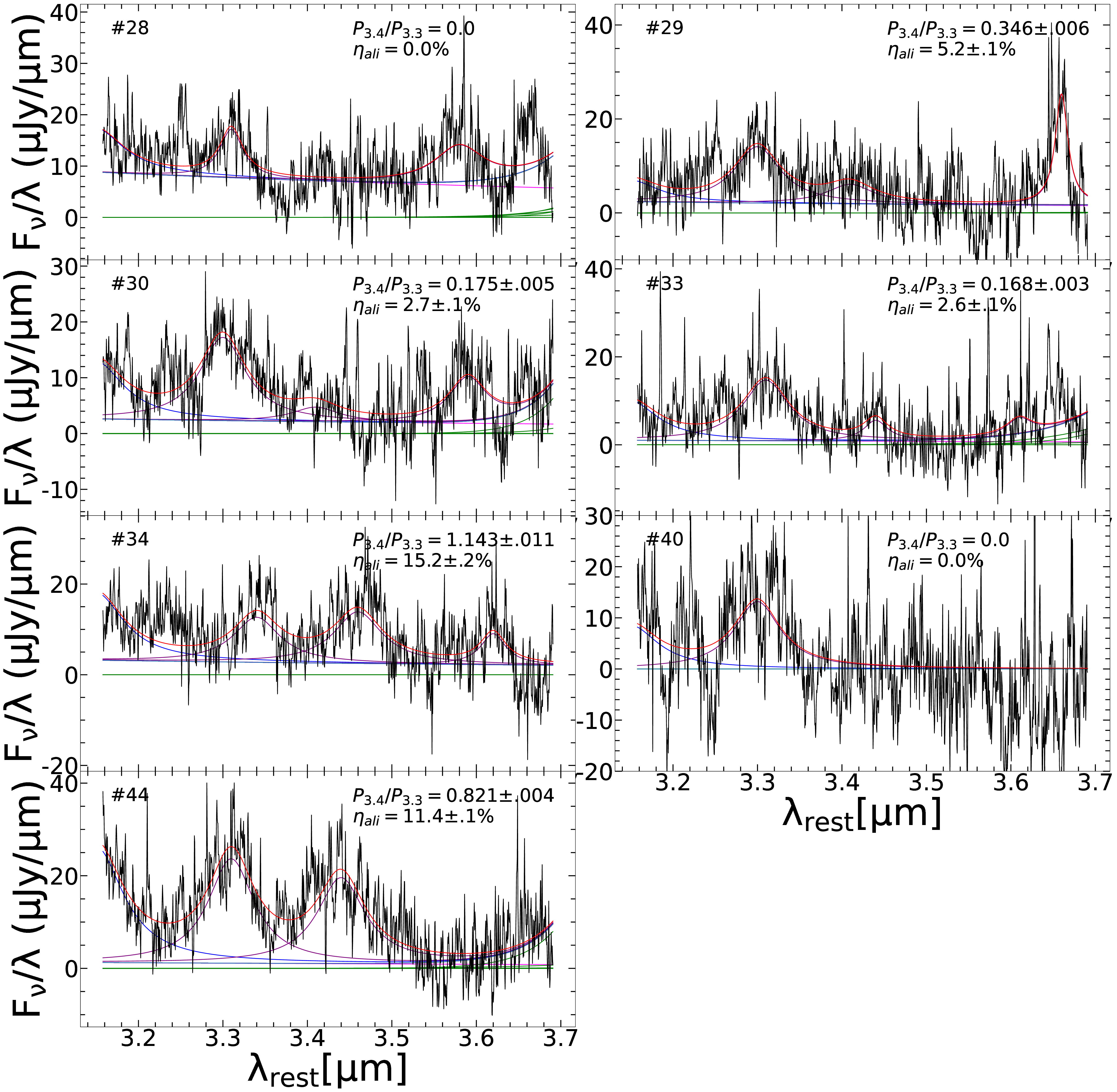}
\end{center}
\vspace{-4mm}
\caption{\footnotesize
  Continued, but for the outermost band.
  }
\vspace{-3mm}
\end{figure*}

\section{Discussion}\label{sec:discussion}
To quantify the intensities of the 3.3 and 3.4$\mum$
emission features, we apply the PAHFIT fitting tool
(Smith et al.\ 2007) to decompose an observed spectrum
into  two Drude profiles and an underlying continuum
(see Figure~\ref{fig:spectra_inner} and
 Figure~\ref{fig:spectra_outer}a,\,b,\,c).
The two Drude profiles characterize the 3.3$\mum$
aromatic and 3.4$\mum$ aliphatic features, respectively.
(In some cases, secondary subfeatures
at $\simali$3.50 and 3.61$\mum$ are also included.)
This allows us to determine the power emitted from
each feature. 


Let $P_{3.3}$ and $P_{3.4}$ be the power emitted
from the 3.3 and 3.4$\mum$ features, respectively.
Let $\alifrac\equiv \left(1+\NCaro/\NCali\right)^{-1}$
be the aliphatic fraction of PAHs, i.e., the ratio of
the number of C atoms in aliphatic units
($N_{\rm C,ali}$) to that in aromatic rings
($N_{\rm C,aro}$) {\it plus} that in aliphatic units.
Yang \& Li (2023) found that $\NCali/\NCaro$
can be derived from the observed  feature ratios
$\Iratioobs$:
\begin{equation}\label{eq:alifrac2}
\frac{\NCali}{\NCaro}
\approx \frac{1}{6.40}
\left(\frac{P_{3.4}}{P_{3.3}}\right)_{\rm obs}~~.
\end{equation}

\begin{figure*}[htp]
\vspace{-1mm}
\begin{center}
\includegraphics[width=12cm,angle=0]{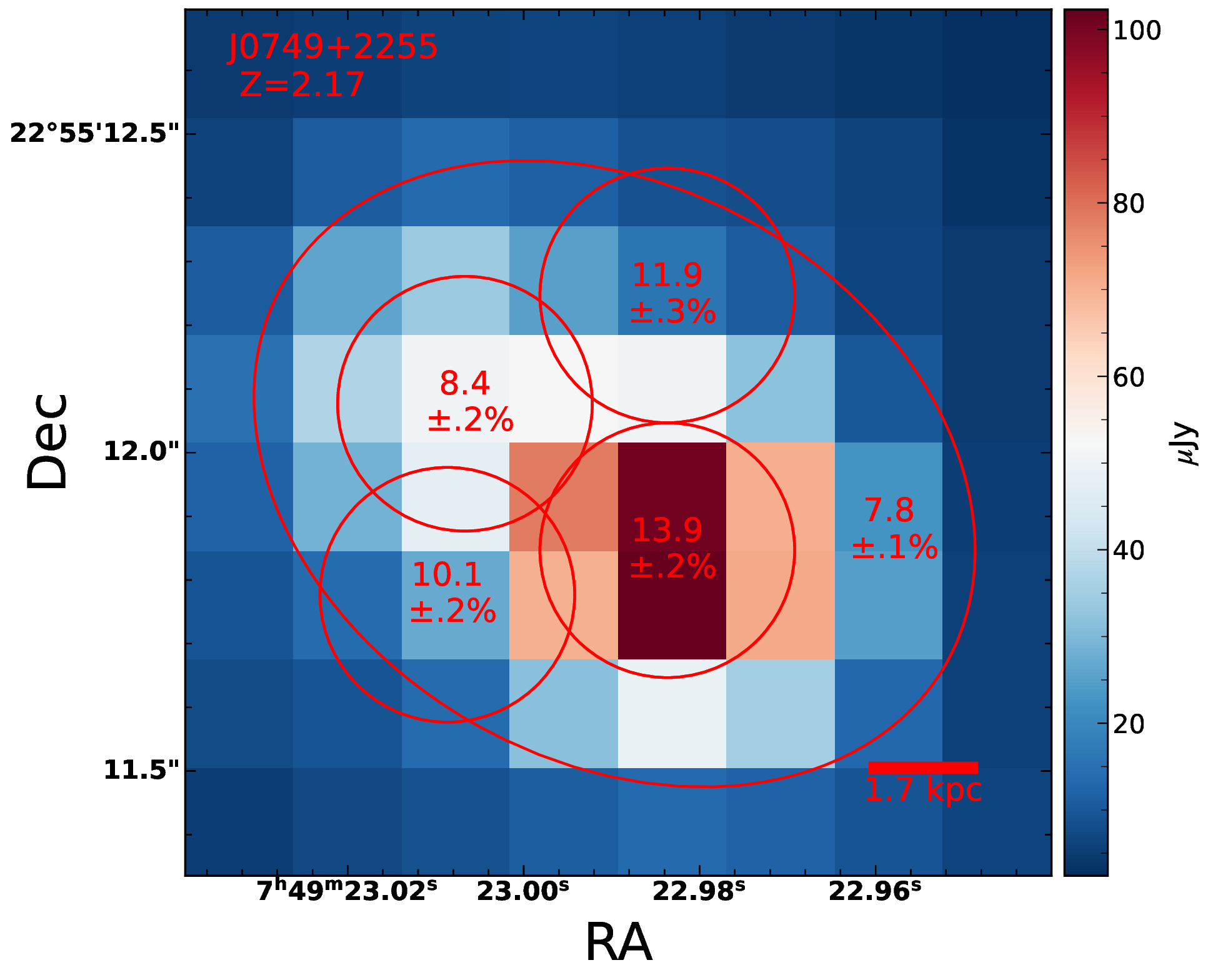}\\
\includegraphics[width=12cm,angle=0]{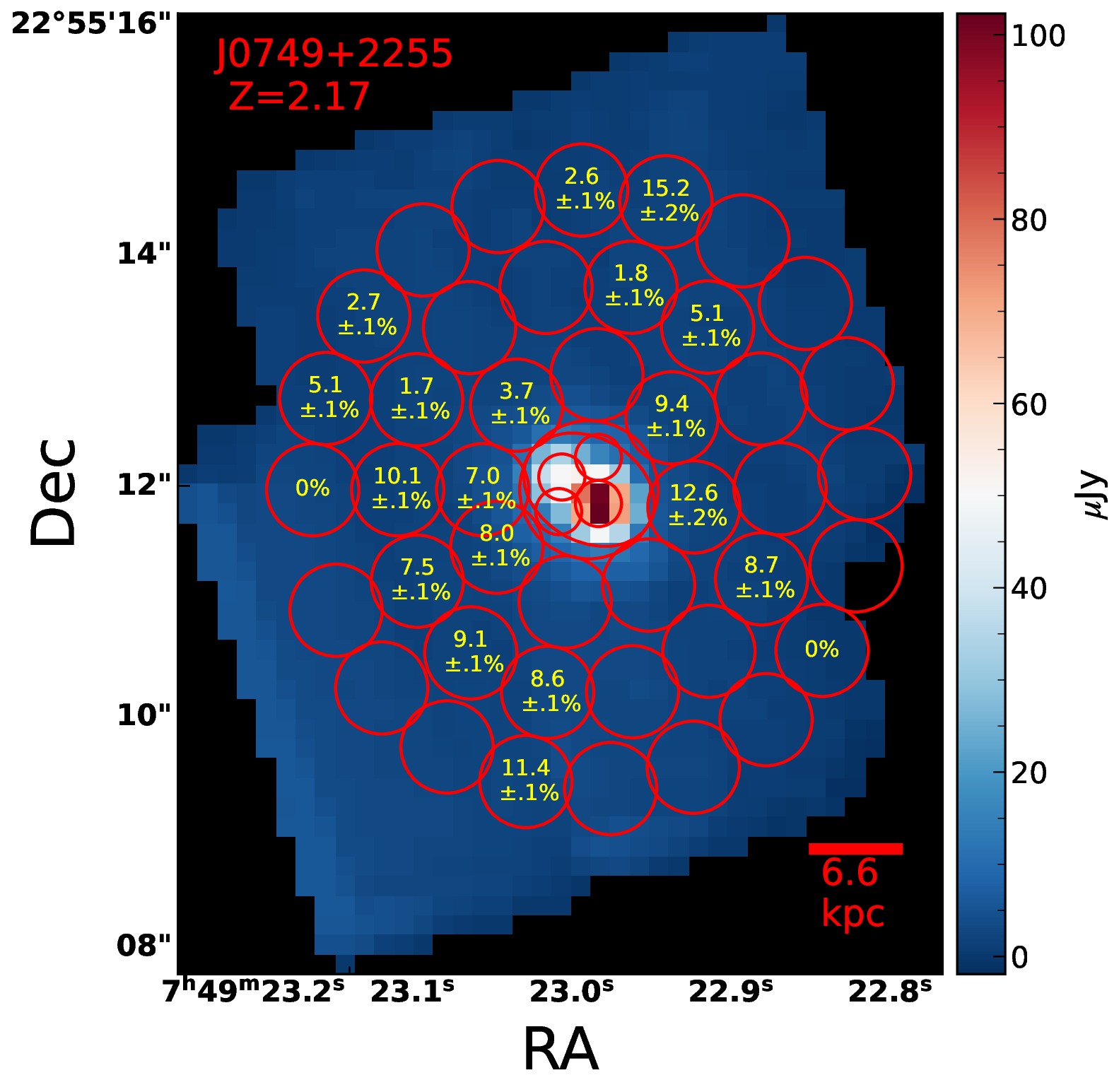}
\end{center}
\vspace{-4mm}
\caption{\label{fig:alifrac}\footnotesize 
  Aliphatic fractions of PAHs
  in the brightest center region (a)
  and the outer region surrounding the center  (b).
  }
\vspace{-3mm}
\end{figure*}

With $P_{3.3}$ and $P_{3.4}$ determined 
from PAHFIT for the JWST/MIRI spectrum
extracted for each aperture, we determine
the PAH aliphatic fraction $\alifrac$ and
list in Figure~\ref{fig:alifrac}.
Clearly, $\alifrac$ exhibits considerable
regional variations.
In the central region, the aliphatic fractions
are remarkably high, from $\simali$8\%
to $\simali$14\%, particularly in one of
the quasar cores (\#0), which is significantly
more luminous than all the other regions.
In the surrounding region,
the aliphatic fractions are lower and show
a greater regional variability,
with only four (of 20) apertures having
$\alifrac>10\%$.

In the brightest central region, particularly
in the two apertures centering on the quasar cores,
the intense UV radiation is expected to
break up the aliphatic sidegroups attached
to PAHs and even destroy small PAHs which
emit at 3.3$\mum$. Therefore, one would not
expect strong 3.3$\mum$ emission or high
aliphatic fractions. It is thus puzzling that
the central region exhibits strong 3.3$\mum$
emission and high $\alifrac$ fractions.
We argue that an aperture of 1.7$\kpc$ is quite large,
even extremely energetic quasars may not be able
to process the interstellar materials on such
a large scale. Sturm et al.\ (2000) obtained
the near- and mid-IR spectra of several nearby
galaxies including Circinus, M82, NGC\,253,
NGC\,1068, and the 30 Doradus stra-forming
complex in the Large Magellanic Cloud (LMC), 
utilizing the {\it Short Wavelength Spectrometer}
(SWS) on board the {\it Infrared Space Observatory}
(ISO). They found that no PAH emission was detected
in NGC\,1068, a prototypical Seyfert 2 galaxy,
as the ISO/SWS observations
were centered on the active nucleus,
with the aperture covering very little of
the circumnuclear star forming regions
(also see Siebenmorgen et al.\ 2004).
In contrast, in the ISO/SWS observations
of Circinus, another Seyfert 2 galaxy,
which were also centered on the active nucleus,
but where the apertures covered a significant
amount of the circumnuclear star formation,
PAH emission was clearly detected.
For Circinus, M82, NGC\,253, and 30 Dor,
the ISO/SWS spectra have an aperture
corresponding to $\simali$0.3--0.4$\kpc$,
which is more or less comparable to 
the $\simali$1.7$\kpc$ scale of
the JWST/MIRI observations of J0749+2255.
For comparison, we overlay the ISO/SWS
spectra of M82 and 30 Doradus
with the JWST/MIRI spectra of \#1
and \#10, respectively.
This demonstrates that on such scales
PAHs survive and their excitation are
dominated by starburst,
rather than AGN-driven shocks.

It is interesting to note that,
Chen et al.\ (2024) analyzed the spatially
resolved [FeII] 5.34$\mum$ emission
(which traces shocks) of J0749+2255,
also obtained with JWST/MIRI.
They found no observable increase
in the 5.34$\mum$ to 3.3$\mum$ [FeII]/PAH ratios
at both cores, as would be anticipated from
enhanced [FeII] emission because of quasar-driven
shocks (Hill \& Zakamska 2014) and/or the disruption
of small PAH molecules by the strong radiation and
shocks from quasars (e.g., see Zhang et al.\ 2023).
This suggests neither the intense quasar UV radiation
nor the quasar-driven outflows (which generate shocks)
penetrate sufficiently into the gas-rich host,
and therefore there is no detectable spatially
resolved sign of either radiative suppression of PAHs 
or quasar-driven shocks that would enhance [FeII] emission. 
It is well recognized that shocks can destroy grains, 
causing iron atoms in dust grain to be sputtered into
the gas phase and subsequently ionized by UV photons
(Mouri \& Taniguchi 2000).
%
%
%
On the other hand, we should also note that
while small PAHs are more easily destroyed
in harsh environments with hot gas and
strong radiation fields (e.g., close to a quasar core),
harder radiation would actually heat PAHs
to higher temperatures, shift their emission
to shorter wavelengths and therefore enhance
the 3.3$\mum$ emission without a significant
change in the small-PAH fraction
(e.g., see Draine et al.\ 2021;
Yadav et al.\ 2026; B.~Yang et al., in preparation).

More recently, based on JWST/NIRSpec observations
of the luminous IR galaxy and dual AGN NGC\,6240,
Carlsen et al.\ (2026) also detected the 3.3$\mum$
PAH emission in the central $\simali$2$\kpc$
of this system and found that the PAH emission
peaks at the two nuclei,
with a morphology differing from the H$_2$ gas.
In this dual AGN system, Carlsen et al.\ (2026)
identified a bi-conical wind launched
from the northern nucleus, as well as an outflow
launched from the southern nucleus,
demonstrating that small PAHs can survive
in shocks induced by high velocity outflows.
Costa-Souza et al.\ (2026) analyzed
the high-resolution JWST/NIRSpec spectra
of the central, circumnuclear regions
of local AGN hosts at $z$\,$\simlt$\,0.1
and detected the 3.3 and 3.4$\mum$
emission in these hostile regions rich in
hard photons and shocks
(e.g., see their Figure~D1
for the 3.3 and 3.4$\mum$ emission
features in the main nucleus of NGC\,3256).
The PAH emission at 3.3 and 3.4$\mum$
as well as those features at longer wavelengths
has also been detected both in the disk
and in the superwinds
of the starburst galaxy M82,
extending $\simgt$5\,kpc from the plane
(Engelbracht et al.\ 2006;
Yamagishi et al.\ 2012;
Beir\~ao et al.\ 2015;
Cronin et al.\ 2026;
J.W.~Lyu et al., in preparation).
Just like J0749+2255,
it is puzzling that the PAH aliphatic fraction is
considerably higher in the superwind than in the center,
as one would imagine that in the harsh superwind
environment PAHs would be easily stripped off
any aliphatic sidegroups and therefore one would
not expect to see strong aliphatic C--H emission
in the superwind.

In the ``outer region'' surrounding the brightest
center, notably, the aliphatic fractions appear to
exhibit a general decrease with increasing distance
from the center. Again, this is intriguing as it contrasts
with the anticipated trend of higher aliphatic fractions
with increasing distance from the center
(as the influence from the UV radiation
and shocks of quasars weaken).
This anomaly suggests that the nature and origin of
the aliphatic emission and the energy profile
of the surrounding environment are more intricate than
previously understood, necessitating further investigation
to elucidate the behavior of aliphatics in highly energetic
contexts and their spatial evolution.

\begin{figure*}[htp]
\vspace{-1mm}
\begin{center}
\includegraphics[width=6.8cm,angle=0]{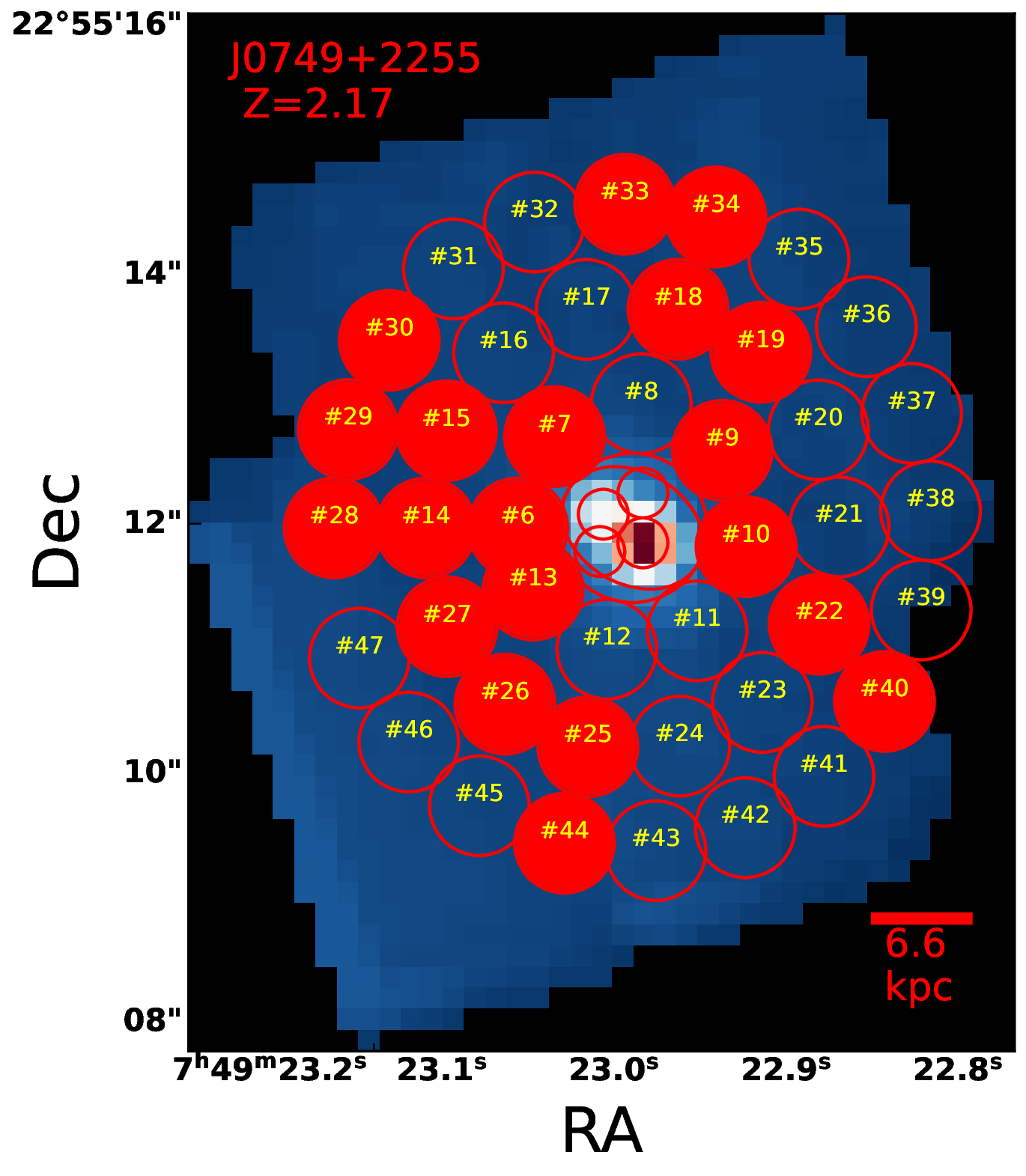}
\includegraphics[width=9.2cm,angle=0]{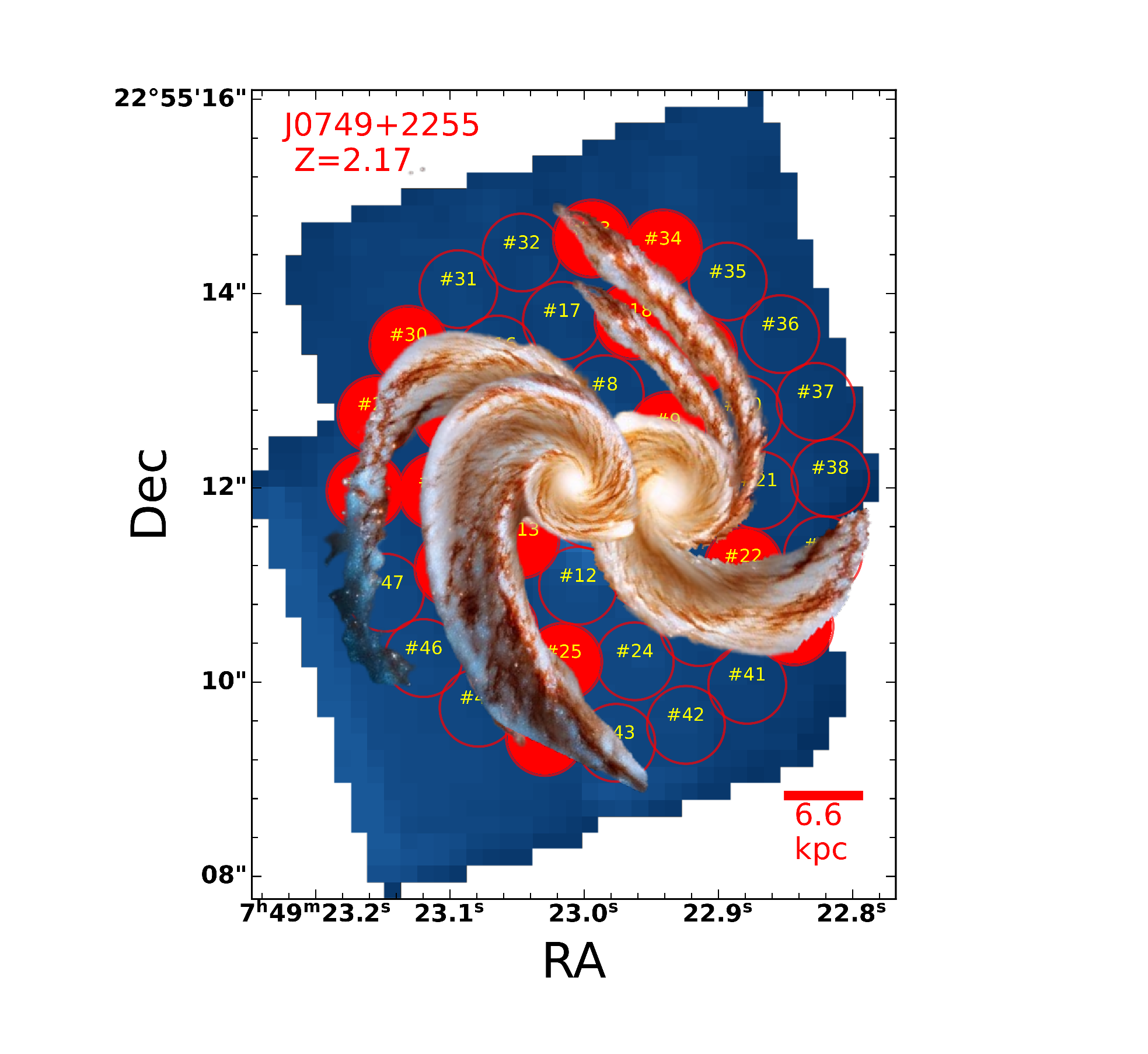}
\end{center}
\vspace{-4mm}
\caption{\label{fig:morph}\footnotesize
Left panel: Apertures with detected 3.3 and 3.4$\mum$
emission highlighted for illustrative purposes.
Right panel: a schematic illustration of potential
hidden structures revealed by the 3.3 and 3.4$\mum$ emission.
}
\vspace{-3mm}
\end{figure*}

Finally, we note that, beyond the cores,
our analysis of the PAH emission reveals
unanticipated structures that were previously
undetectable in the HST/WFC3 or JWST's MIRI/MRS images.
Figure~\ref{fig:morph} illustrates apertures
with detected 3.3 and 3.4$\mum$ emission,
marked in red, revealing two distinct patterns
on opposite sides of the cores.
Recent studies indicate a higher prevalence of
spiral structures at these redshifts 
suggesting that these patterns may represent
trailing arms or tidal tails formed
during the merging process.
An additional schematic illustration
has been provided to hypothesize
the hidden structures potentially
revealed by the 3.3 and 3.4$\mum$ emission,
excited by the starlight emitted from young
stars formed in the tidal tails.
Indeed, faint tidal tails are seen in the deep
HST near-IR images of J0749+2255
(Chen et al.\ 2023).
Recent IFU observations made with
JWST's {\it Near Infrared Spectrograh} (NIRSpec)
revealed a giant rotating disk
perpendicular to the direction
of the two quasar cores, rather than
a disturbed system with irregular merger
morphology (Ishikawa et al.\ 2025).
The 3.3$\mum$ emission could be tracing
this structure through its excitation by UV
photons provided by star formation.
Chen et al.\ (2024) derived from
the 3.3$\mum$ emission an extremely high
star formation rate for J0749+2255,
suggesting that this system is undergoing
intense starburst activities.
It is interesting to note that,
based on JWST's {\it Near Infrared Camera}
(NIRCam) observations of the nine most striking 
ram pressure stripping (RPS) galaxies
in the Abell 2744 cluster at $z$\,$\approx$\,0.306,
Benotto et al.\ (2026) detected the 3.3$\mum$
PAH emission in eight of the nine galaxies,
with morphologies revealing disk truncation
and elongation along the RPS direction.
The 3.3$\mum$ PAH emission is prominent
in star-forming clumps embedded
in the stripped tails up to a distance
of $\simali$40$\kpc$.
They found that the spatial correlation between
the PAH emission, stellar age, and star formation
rate is consistent across disks and tails,
demonstrating that small PAH molecules
can survive and become stripped by ram pressure.


%
\section{Summary}\label{sec:summary}
We have examined the JWST/MIRI IFU data of
the dual quasar system J0749+2255 at $z\approx2.17$.
Our major results are as follows:
\begin{enumerate}
\item The spatially resolved IFU data revealed a widespread
  detection of aromatic and aliphatic C--H emission
  at 3.3 and 3.4$\mum$, marking the highest redshift
  detection of both aromatics and aliphatics in space.
\item We have determined the aliphatic fraction $\alifrac$
  of the carriers of the 3.3 and 3.4$\mum$ emission
  (i.e., small PAHs of several tens of carbon atoms)
  and explored its spatial variations across the system,
  revealing a somewhat decreasing trend for $\alifrac$
  with distance from the quasar cores.
  This contrasts with conventional wisdom
  and highlights the complex dynamics of aromatics
  and aliphatics in highly energetic environments,
  emphasizing the need for further studies to
  understand the behavior of aromatics and aliphatics
  in such extreme conditions.
%
\item The detection of PAH emission in the quasar-surrounding
  environments unveils complex structures
  that are potentially indicative of trailing arms
  and tidal tails formed during the merger process,
  offering insights into the morphology and evolution
  of merging galaxies at high redshifts.
\end{enumerate}
%


\acknowledgments
We thank Y.~C.~Guo, J.W.~Lyu, J.H.~Peng,
and B.~Yang for stimulating discussions.
We thank the anonymous referee
for his/her helpful comments and
suggestions that improved the quality
and presentation of this work.
XJY is supported in part by NSFC\,12333005
and 12122302, CMS-CSST-2021-A09,
and the Innovative Research Group Project of
Natural Science Foundation of Hunan Province
of China No. 2024JJ1008.

\dataset
This work is based on observations
made with the NASA/ESA/CSA
James Webb Space Telescope and
the NASA/ESA Hubble Space Telescope.
The data were obtained from the Mikulski
Archive for Space Telescopes at the Space
Telescope Science Institute, which is operated
by the Association of Universities for Research
in Astronomy, Inc., under NASA contracts
NAS 5–26555 and NAS 5-03127.
The specific observations analyzed
can be accessed via
\dataset[doi:10.17909/3jnh-3w31]
{https://doi.org/10.17909/3jnh-3w31}.


\end{document}